\documentstyle[12pt]{article}
\def\pref#1{\protect{\ref{#1}}}
\newcommand{\mlim}[1]{\,{\rm lim}
                      \hspace*{-5.9ex}\raisebox{-1ex}{$\quad_{#1}$}}
\newcommand{\ch}{{\rm cosh}}
\newcommand{\cth}[1]{{\rm tanh^{-1}}\left(#1\right)}
\newcommand{\sh}{{\rm sinh}}
\newcommand{\arcsh}{{\rm arcsinh}}
\newcommand{\th}{{\rm tanh}}
\newcommand{\arctg}{{\rm arctg}}
\newcommand{\arcth}{{\rm arctanh}}
\newcommand{\const}{{\rm const}}
\newcommand{\ro}{r_{\rm 0}}
\newcommand{\gm}[3]{\Gamma ^#1 _{#2 #3}}
\newcommand{\gf}[3]{\check \Gamma ^#1 _{#2 #3}}
\newcommand{\p}[3]{P ^#1 _{#2 #3}}
\renewcommand{\d}{{\rm d}}
\renewcommand{\exp}[1]{{\rm exp}\left( #1 \right)}
\newcommand{\rf}[2]{\check R _{#1 #2}}

\catcode`\@=11
\newif\if@ABC
\@ABCfalse
\newcommand{\setletter}[1]{
\if@ABC\else\@ABCtrue\global\let\@tMp=\theequation\fi%
           \gdef\theequation{\@tMp #1}%
\ifmmode%
\typeout{**Warning!**  'Setletter' in math mode destroys previous label!}%
\global\let\@currentlabel=\theequation%
\fi%
}
\newcommand{\remletter}{
  \if@ABC\gdef\theequation{\@tMp}%
  \else %
\typeout{**Warning!** Use 'setletter' before 'remletter' !}%
\fi%
             }

\newcommand{\deceqno}{\addtocounter{equation}{-1}}
\global\newcounter{@C@figure}
\newcommand{\CAPT}[1]{%
  \refstepcounter{@C@figure}%
  \relax \noindent {\bf Figure \arabic{@C@figure}.} #1 \par\vskip .5cm\relax%
                         }
\def\tlabel#1{%
   \if@ABC%
{\let\thepage=\relax\edef\next{%
\write \@auxout {\string \newlabel{#1}{{\@tMp}{\thepage}}}%
}\next}%
  \else\typeout{**Warning!** Use 'setletter' before 'tlabel' !}\fi%
                        }
\catcode`\@=12
\begin{document}
\large
\begin{center}
{ \LARGE  Spherically symmetric solutions of gravitation equations
on the background with spatial sections of constant curvature}\\
{M.N.~Tentyukov\\ \normalsize   Joint Institute for Nuclear Research,
Dubna.\\e-mail: tentukov@thsun1.jinr.dubna.su}
\end{center}
\baselineskip 6mm
\vskip 1cm
\begin{sloppypar}
\begin{abstract}
We investigate the vacuum and charged spherically
symmetric static solutions of the Einstein equations on
cosmolo\-gi\-cal background. The background metric
is not flat, but curved, with constant - curvature spatial sections.

Both vacuum and charged cases contain
two branches. The first branches transform
into the Schwarzschild and Reissner-Nordstr\"om solutions
if the background metric goes to the Minkovski one. The
second branches describe wormholes
and have no Einstein limit.
\end{abstract}

\section{Introduction}

It is well - known that the equations describing gravitation with a given
background metric coincide with the Einstein ones if the background metric
satisfies the "background" Einstein equations \cite{Grischuk-Popova}.
However, if the background metric is arbitrary, the gravitational equations
differ from the Einstein equations. Some exact solutions of these equations
which are in form very close to the Einstein solutions acquire new specific
features.

There are at least  two reasons for these solutions to be interesting.

First, they are solutions of the Einstein
equations on the nontrivial cosmological background. That is
why they may be physically meaningful in the early Universe.

When the quantum properties of gravity were important, the Einstein
theory was to be replaced by the effective theory that includes
quantum effects. Later on, gravitation was described by the Einstein
equations.
However, the intermediate period between quantum and Einstein states
may admit the description in terms of the
Einstein gravity on the quantum - corrected background.

Solutions on the nontrivial background may be useful
for the functional integral approach to quantum gravity. In the
semiclassical approximation, the dominant contribution to the
path integral will come from the neighborhood of saddle points of the action,
that is, of classical solutions. It seems to be interesting to compare
the path integral based on the Gibbos - Hawking action with another model,
whose properties may be different.

Of course, we can write down any metric which does not satisfy any
equations. However, to derive the Lagrangian is a more complicated problem.
Indeed, $R$ is the only scalar without essential higher derivatives.
All other Lagrangians contain either higher derivatives or additional
dynamical degrees of freedom.

Introducing a nondynamical background object, we can
derive solutions with very different properties and, at the same time,
retain the basic structure of the Lagrangian.

\section{Basic results}

In the present paper, we consider the gravitation described by the
spherically symmetric static metric
$$
      \d s^2=g_{ik}\d x^i \d x^k
$$
on the background
\setletter{a}
\begin{equation}                          \tlabel{1} \label{1a}
      \d\check s^2=\d t^2-\d r^2-k^2\sh^2\frac{r}{k} \d\Omega,
\end{equation}
\deceqno
where $\d\Omega= \left(\d\theta^2+\sin^2\theta\, \d\phi^2\right)$.
The spatial sections $t=\const$ are 3-d Lobachevski spaces
(a constant negative curvature space of radius $k$).
For simplicity, we use the system of units $\hbar=c=G=k_\beta=1$.
It is convenient to shift a coordinate $r$ to some value
$\ro$ so that (\ref{1a}) takes the form
\setletter{b}
\begin{equation}                                    \label{1b}
      \d\check s^2=\d t^2-\d r^2-k^2\sh^2\frac{r-\ro}{k} \d\Omega.
\end{equation}
\remletter

The equations
\begin{equation}                                 \label{2}
R_{ij}-\rf{i}{j} =8\pi (T_{ij}-\frac{1}{2}g_{ij}T),
\end{equation}
where $\rf{i}{j} $ is the background Ricci tensor, $T_{ij}$
is the energy - momentum tensor of the external matter,
are derived from the action
$$
S=\int(L_g+L_M)\,\d^4x
$$
with the Rosen \cite{Rosen} gravitational Lagrangian
\begin{equation}                                 \label{3}
L_g=\sqrt{-g}g^{mn}(\p{a}{m}{b}\p{b}{a}{n}
- \p{a}{s}{a}\p{s}{m}{n}).
\end{equation}
Here the affine - deformation tensor
$$
\p{i}{j}{k}=\gf{i}{j}{k}-\gm{i}{j}{k}
$$
is the difference between the background connection and the
Levi - Chivita connection for physical metric.
The matter Lagrangian $L_M$ does not contain background objects.

We will investigate the vacuum (\ref{5}) and charged (\ref{8}) spherically
symmetric static solutions of equations (\ref{2}). Both the cases contain
two branches. The first branches (\ref{6}) and (\ref{9}) transform
into the Schwarzschild and Reissner-Nordstr\"om solutions
if we put $k \to \infty$. The
second branches (\ref{7}) and (\ref{12}) describe wormholes
and have no Einstein limit (they diverge when $k \to \infty$).

A charged wormhole can contain an event horizon and the Cauchy horizon in
the throat. Then structure of the whole geodesic space - time
is similar to the Kerr solution.
It is interesting that the  corresponding black holes have
a positive heat capacity.

Equations (\ref{2}) on the background (\ref{1}) were considered for a
localized mass \cite{ChernActaPhPol24} and the electric charge
\cite{Asanov}.
If we put $k \to \infty$, the metric (\ref{1}) goes over to the
Minkowski metric, and equations (\ref{2}) turne into the Einstein
equations.

Arbitrary spherically symmetric static metric may be written
in the form
\begin{equation}                                             \label{4}
\d s^2=\Lambda (r)\d t^2-\frac{1}{\Lambda(r)}\d r^2 -R^2\d\Omega^2.
\end{equation}
The causal structure of space - time is determined by the
metric function $\Lambda$, and $R(r)$ is the "luminosity" distance.
It may be shown that, if $R(r)$ is an increasing function, the
decreasing function $\Lambda$ describes repulsion of a test body
from the central source; while increasing $\Lambda$, attraction.

Let us denote
\begin{eqnarray}
      \Lambda_1&=&\exp{2\ro/k}
      \frac{\sh\frac{r-2\ro}{k}}{\sh\frac{\,r\,}{k}},
      \nonumber\\
      \Lambda_2&=&\exp{2\ro/k}
      \frac{\ch\frac{r-2\ro}{k}}{\ch\frac{\,r\,}{k}},
      \nonumber\\
      R_1&=&\exp{-\ro/k}k\,\sh\frac{r}{k},
      \nonumber\\
      R_2&=&\exp{-\ro/k}k\,\ch\frac{r}{k}.
      \nonumber
\end{eqnarray}
First of all we consider the vacuum case $T_{ij}=0$:
\begin{equation}                                        \label{5}
R_{ij}=\rf{i}{j}.
\end{equation}

Besides the solution
\begin{equation}                                        \label{6}
\d s^2_{\rm V1}=\Lambda_1\d t^2 - \frac{1}{\Lambda_1}\d r^2 -
R^2_1d\Omega^2
\end {equation}
transforming to the Schwarzschild solution when $k \to \infty$
there is the second solution of (\ref{5})
\begin{equation}                                        \label{7}
\d s^2_{\rm V2}=\Lambda_2\d t^2 - \frac{1}{\Lambda_2}\d r^2 -
R^2_2\d\Omega^2
\end{equation}
describing traversible wormhole.
This second branch demands nontrivial background topology
continued to negative $r$.

The solution (\ref{6}) was found in \cite{ChernActaPhPol24}.
It contains a Schwarzschild - like singularity
on the event horizon at $r=2\ro$.
The metric function and the luminosity distance are shown in fig.\ref{fig1}.

We may find the coordinate system in which (\ref{6}) is regular at
$r=2\ro$ but  in these coordinates there is the background metric
(\ref{1}) singularity. Since the background metric is
non - observable, we are interested only in the physical metric
(\ref{6}). The causal structure of maximal analytic extension (\ref{6})
is the same as the Schwarzschild one.
Note that if $\ro \ll k$, the value of $\ro$ is very close
to the value of  mass of the central source. If $\ro \to 0$, then
(\ref{6}) goes over to the background metric (\ref{1}). The asymptotics
of (\ref{6}) for $r \to \infty$ coincides with the asymptotics of
(\ref{1b}).

The metric function of (\ref{7}) is regular at all $r$
and the luminosity distance has a minimum
at $r=0$ (see fig.\ref{fig2}). It means that if
$r$ is smaller than $\ro$, the metric (\ref{7}) describes a
traversible wormhole. The locally defined background (\ref{1})
derives locally the physical metric (\ref{7}) by means of
equations (\ref{5}). The geodesic completeness  of (\ref{7})
leads to the topology both of background and physical metrics.

Solution (\ref{7}) diverges if $k \to \infty$ because the
minimum of $R_2(r)$ goes to $\infty$. Asymptotics of (\ref{7})
for $r \to \infty$ is the same as that of (\ref{6}) and (\ref{1b})
but if $ \ro \to 0$, the metric (\ref{7}) turns to the metric
$$
\d s^2=\d t^2-\d r^2-k^2 \ch^2\frac{r}{k}\d\Omega^2
$$
describing a symmetric wormhole in the absolute empty space.

If $\ro > 0$, the asymptotics  of the physical metric
for $r \to -\infty$ differs from the background one.
It can be shown that the wormhole sucks in  a matter from the 3-space
with a nonbackground asymptotics and then throws out it into the 3-space
whose asymptotics is the same as the background one. However, a rapidly
moving radial observer can visit another 3-space and then return to his
home during a finite proper time.

Now let us consider equations (\ref{2}) with the energy - momentum tensor
of a localized electric charge $Q$:
\begin{equation}                                 \label{8}
R_{ij}-\rf{i}{j} =8\pi T_{{\rm Q}ij}.
\end{equation}
Here we also obtain two branches. The first
\begin{equation}                                        \label{9}
\d s^2_{\rm Q1}=\Lambda_{\rm Q1}\d t^2 - \frac{1}{\Lambda_{\rm Q1}}\d r^2 -
R^2_1d\Omega^2,
\end{equation}
where $$
   \Lambda_{\rm Q1}= \Lambda_1 +\exp{\!\frac{4\ro}{k}\!}\frac{Q^2}{k^2}\,
   \sh^{-2}\frac{r}{k}
   $$
was found in \cite{Asanov}.
When $k \to \infty $, it passes into the Reissner-Nordstr\"om solution.
If $Q=0$, then (\ref{9}) coincides with (\ref{6}). If $Q<Q_{10}$,
where
\begin{equation}                                                  \label{10}
  Q_{10}=\frac{k\,\sh\frac{\ro}{k}}{\sh\frac{\ro}{k}+\ch\frac{\ro}{k}},
\end{equation}
there are two horizons at $r_{1+}$ and $r_{1-}$,
\begin{equation}                                                  \label{11}
r_{1\pm}=\ro \pm k\,\arcsh\sqrt{\sh^2\frac{\ro}{k}-\frac{Q^2}{k^2}\left(
\sh\frac{\ro}{k}+\ch\frac{\ro}{k}\right)^2}.
\end{equation}

When $Q>Q_{10}$, horizons are absent. If $Q=Q_{10}$, the horizons
merge into a single horizon at $r=\ro$ and an extreme black hole forms.

The second branch
\begin{equation}                                        \label{12}
\d s^2_{\rm Q2}=\Lambda_{\rm Q2}\d t^2 - \frac{1}{\Lambda_{\rm Q2}}\d r^2 -
R^2_2\d\Omega^2,
\end{equation}
where $$
   \Lambda_{\rm Q2}= \Lambda_2 -\exp{\!\frac{4\ro}{k}\!}\frac{Q^2}{k^2}
   \ch^{-2}\frac{r}{k},
   $$
should be interpreted similarly to the metric (\ref{7}).

If $Q=0$, then (\ref{9}) coincides with (\ref{7}). If $Q<Q_{20}$,
where
\begin{equation}                                                  \label{13}
  Q_{20}=\frac{k\,\ch\frac{\ro}{k}}{\sh\frac{\ro}{k}+\ch\frac{\ro}{k}},
\end{equation}
metric (\ref{12}) describes the traversible wormhole.
If $Q=Q_{20}$, there is a horizon in the throat.
When $Q>Q_{10}$, there are two horizons at $r_{2+}$ and $r_{2-}$,
\begin{equation}                                                  \label{14}
r_{2\pm}=\ro \pm k\,\arcsh\sqrt{-\ch^2\frac{\ro}{k}+\frac{Q^2}{k^2}\left(
\sh\frac{\ro}{k}+\ch\frac{\ro}{k}\right)^2}.
\end{equation}
Note that from (\ref{10}) and (\ref{13}) we have
$$Q_{10}=Q_{20}\th\frac{\ro}{k}.$$ Since $\th\frac{\ro}{k}<1$, we have
$Q_{10}<Q_{20}$. Because in (\ref{9}) a horizon exists if $Q<Q_{10}$,
and in (\ref{12}) it exists if $Q>Q_{20}$, for any $\ro$ there is an
area of $Q$ in which there are horizons neither in (\ref{9}) nor
in (\ref{12}).

Now we return to solution (\ref{12}).

When $Q=\mlim{\ro\to 0}Q_{20}=k$, horizons depend on $\ro$ as
shown in fig.\ref{fig3}. If $Q<k$, for small enough
$\ro$ a horizon is absent. If $Q>k$, then even for $\ro = 0$ there are two
horizons at $r_{2\pm \rm min}=\pm k\,\arcsh\sqrt{\frac{Q^2}{k^2}-1}$
(fig.\ref{fig4}).

When there are two horizons, the causal structure of (\ref{12})
is similar to the Kerr solution but without singularity. The conformal
diagram is shown in fig.\ref{fig5}.

Temperature of the solutions with horizons can be found by investigating the
Euclidean sections of the corresponding metrics \cite{GibHaw}.

For (\ref{6}) and (\ref{9}) the corresponding temperatures are
\begin{eqnarray}                                                    \label{15}
T_{\rm V1}&=&\frac{\exp{\frac{2\ro}{k}}}{4\pi k}
            \frac{1}{\sh\frac{2\ro}{k}}=
                   \frac{1}{4\pi k}\left( 1+\cth{\frac{2\ro}{k}} \right) \\
                                                                    \label{16}
T_{\rm Q1}&=&\frac{\exp{\frac{2\ro}{k}}}{4\pi k}
            \frac{\sh\,\frac{r_{1+} - r_{1-}}{k}}
                 {\sh^2\frac{\,r_{1+}\,}{k}}
\end{eqnarray}

The dependence of $T_{\rm Q1}$ on $\ro$ is very close to the dependence
of $T_{\rm V1}$, if $\ro$ is not very small. If $k \to \infty$,
then (\ref{15}) and (\ref{16}) pass into
expressions for temperatures of the Schwarzschild
and Reissner-Nordstr\"om solution, respectively. When $\ro \to 0$, then
$T_{\rm V1} \to \infty$ as the Schwarzschild temperature. The behavior of
$T_{\rm V1}$ is similar, but when $\ro$ is very small, then for any
(small) $Q$ the horizon does not form, and at $Q=Q_{10}$ the temperature
$T_{\rm Q1} =0$. That is why for any $Q\neq 0$ there is a neighborhood of $0$
for $\ro$ in which $T_{\rm V1}$ and $T_{\rm Q1}$ depend on $\ro$
as shown in fig.\ref{fig6}.

For large $\ro$ the situation differs from the GR one:
$$\mlim{\ro \to \infty}T_{\rm V1}=\mlim{\ro \to \infty}T_{\rm Q1}=
  \frac{1}{2\pi k}.
$$
It means that very heavy black holes can't have temperature less than
$\frac{1}{2\pi k}$.

The solution (\ref{12}) has two temperatures.
In the 3-space when $r_{2+}$ is the
event horizon and $r_{2-}$ is the Cauchy horizon, we observe
the temperature
\begin{equation}                                                \label{17}
T_{\rm Q2+}=\frac{\exp{\frac{2\ro}{k}}}{4\pi k}
              \frac{\sh\,\frac{r_{2+}-r_{2-}}{k}}
                   {\ch^2\frac{\,r_{2+}\,}{k}},
\end{equation}
and in the 3-space, when $r_{2-}$ is the
event horizon and $r_{2+}$ is the Cauchy horizon, we observe
the temperature
\begin{equation}                                                \label{18}
T_{\rm Q2-}=\frac{1}{4\pi k}
              \frac{\sh\,\frac{r_{2+}-r_{2-}}{k}}
                   {\ch^2\frac{\,r_{2-}\,}{k}}.
\end{equation}
These temperatures are increasing functions of $\ro$! Therefore,
the heat capacity
of (\ref{12}) is positive definit, and so we can use a normal canonical
ensemble.
The lower limit for both the temperatures is equal to
\begin{equation}                                              \label{19}
l=\frac{\sqrt{1-\frac{k^2}{Q^2}}}{2\pi k}.
\end{equation}
The maximal value of $T_{\rm Q2+}$ is $\frac{1}{2\pi k}$, and
the maximal value of $T_{\rm Q2-}$ is
\begin{equation}                                              \label{20}
L=\frac{
        \left(
              1-4\frac{Q^2}{k^2}
        \right)^2
       }
       {
        32 \pi k\frac{Q^4}{k^4}
       }
\end{equation}
{}From (\ref{19}) we can see that for $Q=k$ the minimal of temperature is
zero, and if $Q>k$, then the minimal temperature is large than zero.
When $Q<k$, then horizons will form only if $\ro$ is not very small,
and the lower temperature corresponds to the minimal value of $\ro$
obtained from (\ref{13}):
$$
r_{0 \, \rm min}= k\,\arcth\frac{k-Q}{Q}.
$$

\section{Causal structure of the vacuum solution}

Let us consider the common metric (\ref{4}) and go to the coordinates
$U,V$:
$$
\left\{
\begin{array}{rcl}
        t+f(r)&=&U,\\
        t-f(r)&=&V.
\end{array}
\right.
$$
If $\Lambda f^{\prime 2} = \frac{1}{\Lambda}$, that is
\begin{equation}                                                   \label{21}
f=\pm\int\frac{\d r}{\Lambda},
\end{equation}
then $U$ and$ V$ are null coordinates and the
metric (\ref{4}) takes the form
$$ \d s^2 = \Lambda \d U \d V + R^2 \d\Omega^2.$$
We are interested in the case when $\d\Omega =0$, so,
\begin{equation}                                                  \label{22}
\d s^2 = \Lambda \d U \d V.
\end{equation}

At a horizon we have $\Lambda = 0$. To avoid this peculiarity,
we can attempt to perform the coordinate transformation
$ \tilde U=\tilde U( U),\quad \tilde V=\tilde V(V)$
so that
\begin{equation}                                                   \label{23}
\d s^2 = \frac{
               \Lambda
              }
              {
                \frac{\d\tilde U}{\d U}\frac{\d\tilde V}{\d V}
              }
                \d\tilde U\d\tilde V.
\end{equation}
If $\frac{\d\tilde U}{\d U}\frac{\d\tilde V}{\d V}$ at a horizon
has the same peculiarity as $\Lambda$, then the metric (\ref{23}) becomes
regular.

For the metric (\ref{6}) we have
$$ f=A\left( Br+\ln\left|\sh\frac{r-2\ro}{k}\right|\right),$$
where $ B=\frac{1}{k}\cth{\frac{2\ro}{k}},\quad
A=\frac{k}{1+kB}$.
In the coordinates
\begin{equation}                                                   \label{24}
\left\{
\begin{array}{rcl}
        \tilde U&=&\exp{\frac{U}{2A}},\\
        \tilde V&=&-\exp{-\frac{V}{2A}},
\end{array}
\right.
\end{equation}
we get the metric(\ref{6}) in the form (\ref{23}):
$$
\d s^2_{\rm V1}=4A^2\exp{\!\frac{2\ro}{k}\!}
\frac{
        1
     }
     {
        \sh\frac{r}{k}\,\,\exp{Br}
     }
\d\tilde U\d\tilde V.
$$.
The conformal diagram constructed similarly to the Schwarzschild one.

\section{The wormhole}

Although the coordinate $r$ in (\ref{1b}) changes only in the
area $\ro < r < \infty$,
we can formally consider continuation of the functions $\Lambda(r)$ and
$R(r)$ onto all the real numbers $-\infty < r < \infty$.

Now let us consider the solution (\ref{7}).

As we can see from fig.\ref{fig2}, the metric (\ref{7})
describes gravitational repulsion if $r>0$. The space - time described
by (\ref{7}) is regular at all $r$. Test particles can cross the sphere
$r=0$ and reach negative values of $r$. Actually, (\ref{7}) contains
the traversible wormhole at $r=0$ connecting two infinitely 3-spaces
$r<\ro$ and $r>\ro$. For this interpretation, we must extend the
background (\ref{1b}) to the area $r<\ro$ (or (\ref{1a}) to $r<0$)
and consider
the background space as two the Lobachevsky spaces glued at $r=0$.

On this background, the solution
(\ref{7}) of equations (\ref{5}) is defined at $-\infty<r<\infty$.
For investigation of the region with $r<\ro$, we can do the
following coordinate and constant transformations:
\begin{equation}                                             \label{25}
\begin{array}{rclrcl}
      \tilde r&=&-r\,\,\exp{-\frac{2\ro}{k}};&\quad
      \tilde t&=&t\,\,\exp{\frac{2\ro}{k}};\\
      \tilde k&=&k\,\,\exp{-\frac{2\ro}{k}};&\quad
      \tilde \ro&=&\ro\,\exp{-\frac{2\ro}{k}}.
\end{array}
\end{equation}
Note that $\tilde \ro/\tilde k=
\ro/k, \quad k>\tilde k$.
The metric (\ref{7}) takes the form
\begin{equation}                                             \label{26}
\d \tilde s^2_{\rm V2}=
\tilde\Lambda_2\d \tilde t^2 - \frac{1}{\tilde\Lambda_2}\d \tilde r^2 -
\tilde R^2_2\d\Omega^2,
\end {equation}
where $$
\tilde\Lambda_2=\exp{\!-\frac{2\tilde \ro}{\tilde k}\!}\frac{\ch
\frac{\tilde r+2\tilde \ro}{\tilde k}}{\ch
\frac{\,\,\tilde r\,\,}{\tilde k}},\quad
\tilde R_2 = \exp{\tilde \ro/\tilde k}\tilde k \,\ch \frac{\tilde r}{\tilde k}.
$$

Asymptotics (\ref{26}) for $\tilde r \to \infty$ coincides with
the asymptotics of the Lobachevski space with radius $\tilde k$.
To make this clear, we should pass to the
coordinate $\rho = \tilde r + \tilde \ro$.
If $\tilde r > 0$, the metric (\ref{26}) describes gravitational
attraction. In fig.\ref{fig8} it is shown conditional
representation of the space-time of the wormhole.
In fig.\ref{fig8} we can see the embedding
diagram which is two dimensional model
of the spatial section at a specific moment of time, embedded
into a fictitious 3 - dimensional space.
Note that the background topology differs from the physical one. A continuous
line in the background space corresponds to the discontinuity of
the physical line at $r=\ro$. To avoid this phenomenon, we should
further change the background topology and consider the background 3-space
as a limit when $\varepsilon \to 0$ of two spatial sections (\ref{1b})
glued on the sphere $\ro+\varepsilon$ (fig.\ref{fig7}).

\section{Radial geodesics}

Equations of geodesics can be obtained from the
Lagrangian
$$
L=\frac{1}{2}g_{ij}\dot x^i\dot x^j,
$$
where $\dot x$ means $\frac{\d}{\d \tau}$ and $\tau$ is a proper time
\cite{Chandr}. For radial geodesics in the metric (\ref{4}) this Lagrangian
takes the form
\begin{equation}                                            \label{27}
 L=\frac{1}{2}\left(\Lambda \dot t^2-\frac{1}{\Lambda}\dot r^2\right).
\end{equation}
The Euler equation for (\ref{27}) is
$$\frac{\d}{\d \tau}(\Lambda \dot t)=0.
$$
{}From this formula we get $\dot t^2=\frac{D^2}{\Lambda^2}$, where $D=\const$
is a constant of motion. Let us substitute it into (\ref{27}) and
take into account that $\dot x^i\dot x^j g_{ij}=1$. We obtain
\begin{equation}                                       \label{28}
\Lambda+v^2=D^2,
\end{equation}
where $v=\dot r$. Consequently, the proper time of radial motion from
the point with coordinate $r_1$ to the point with coordinate $r_2$ is
$$\tau_{12}=\pm\int\limits_{r_1}^{r_2} \frac{\d r}{\sqrt{D^2-\Lambda(r)}}.$$
Since direction of motion is towards the center we have for $r_1>r_2$
that $\tau_1<\tau_2$. Finally, the proper time of radial motion from $r_1$
to $r_2$ is
\begin{equation}                                            \label{29}
\Delta\tau=\int\limits_{r_2}^{r_1} \frac{\d r}{\sqrt{D^2-\Lambda}}.
\end{equation}

Let us consider a free falling body in the metric (\ref{9})
with the parabolic
velocity $\mlim{r \to \infty}v=0$. From (\ref{28}) we get $D_1^2=1.$
Let
\begin{equation}                                           \label{30}
\begin{array}{l}
T_1(r)=\int\frac{\d r}{\sqrt{1-\Lambda_1}}=\\
\\
=\frac{
        k
     }
     {
        \sqrt{\exp{\frac{4\ro}{k}}-1}
     }
\left(
      \sqrt{\exp{\frac{2r}{k}}-1}-
      \arctg\sqrt{\exp{\frac{2r}{k}}-1}
\right).
\end{array}
\end{equation}
{}From(\ref{29}) we obtain the proper time of free fall from
$r=r_i$ to $r=r_f<r_i$:
$$\Delta\tau = T_1(r_i)-T_1(r_f).$$
{}From (\ref{30}) we can see that $\Delta\tau$ diverges when $r_i \to \infty$
but it is finite for any other $r_i$ and $r_f$.

Now let us consider the fall in metric (\ref{7}). In this case we cannot
assume $v \to 0 $ when $r \to \infty$ because test bodies are repulsed
by the center. That is why we put $v=0$ at $r=-\infty$:
$$D^2_2=\mlim{r \to -\infty} \Lambda_2 =\exp{\!\frac{4\ro}{k}\!}.$$
{}From(\ref{29}) we obtain the proper time of free fall from
$r=r_i$ to $r=r_f<r_i$:
$$\Delta\tau = T_2(r_i)-T_2(r_f),$$
where
\begin{equation}                                           \label{31}
\begin{array}{l}
T_2(r)=\\
=\frac{
        k
     }
     {
        \sqrt{\exp{\frac{4\ro}{k}}-1}
     }
\left(
      -\sqrt{\exp{-\frac{2r}{k}}+1}-
      \frac{1}{2}\ln
        \frac{
               \sqrt{\exp{-\frac{2r}{k}}+1} -1
             }
             {
               \sqrt{\exp{-\frac{2r}{k}}+1} +1
             }
\right).
\end{array}
\end{equation}
It is clear from (\ref{31}) that $\Delta\tau$ is
finite for all finite $r_i$ and $r_f$
and diverges both for $r_i \to \infty$ ($\ln$) and for $f_f \to -\infty$
(the first term in the brackets).

\section{The metric functions of the charged solution}

Sets of the metric functions (\ref{9}) and (\ref{12}) for different $Q$
are shown in fig.\ref{fig10} and fig.\ref{fig11}, respectively.

If $Q \leq Q_{10}$ (see (\ref{10})), then the metric function $\Lambda_{Q1}$
may be represented in the form
\begin{equation}                                           \label{32}
\Lambda_{Q1}=\exp{\!\frac{2\ro}{k}\!}
\frac{
        \sh\left(\frac{r-r_{1-}}{k}\right)
        \sh\left(\frac{r-r_{1+}}{k}\right)
     }
     {
        \sh^2\frac{r}{k}
     },
\end{equation}
with the horizons
\begin{equation}                                          \label{33}
r_{1\pm}=\ro\pm k\Delta_1.
\end{equation}
Here $\Delta_1$ can be derived by the following formulas:
\begin{equation}                                            \label{34}
\sh\,\Delta_1=
\sqrt{\sh^2\frac{\ro}{k}-\frac{Q^2}{k^2}\left(\sh\frac{\ro}{k}+\ch\frac{\ro}{k}\right)^2}
\end{equation}
or
\begin{equation}                                            \label{35}
\ch\,\Delta_1=
\sqrt{\ch^2\frac{\ro}{k}-\frac{Q^2}{k^2}\left(\sh\frac{\ro}{k}+\ch\frac{\ro}{k}\right)^2}.
\end{equation}

The metric function $\Lambda_{Q2}$ has a similar representation:
\begin{equation}                                           \label{36}
\Lambda_{Q2}=\exp{\!\frac{2\ro}{k}\!}
\frac{
        \sh\left(\frac{r-r_{2-}}{k}\right)
        \sh\left(\frac{r-r_{2+}}{k}\right)
     }
     {
        \sh^2\frac{r}{k}
     },
\end{equation}
where
\begin{equation}                                          \label{37}
r_{2\pm}=\ro\pm k\Delta_2,
\end{equation}
and
\begin{equation}                                            \label{38}
\sh\,\Delta_2=
\sqrt{-\ch^2\frac{\ro}{k}+\frac{Q^2}{k^2}\left(\sh\frac{\ro}{k}+\ch\frac{\ro}{k}\right)^2}
\end{equation}
or
\begin{equation}                                            \label{39}
\ch\,\Delta_2=
\sqrt{-\sh^2\frac{\ro}{k}+\frac{Q^2}{k^2}\left(\sh\frac{\ro}{k}+\ch\frac{\ro}{k}\right)^2}.
\end{equation}

\section{Causal structure of charged solutions}

For transition to the coordinates (\ref{22}) we need the function $f$
(\ref{21}). We are interested in solutions with horizons, therefore,
metric functions may be taken in the form (\ref{32}) and (\ref{36}).
The corresponding integrals can be easily calculated. The forms for both
the functions are the same up to constants:
$$
f=Ar+B\ln\left|\sh\frac{r-r_{-}}{k}\right| +
  C\ln\left|\sh\frac{r-r_{+}}{k}\right|.
$$
For the metric (\ref{9})
\begin{eqnarray}
\nonumber A_1&=&\exp{\!-\frac{2\ro}{k}\!}\ch\frac{r_{1-}+r_{1+}}{k};\quad
          B_1=-\exp{\!-\frac{2\ro}{k}\!}
          \frac{
                  k\,\,\sh^2\frac{r_{1-}}{k}
               }
               {
                  \sh\frac{r_{1+}-r_{1-}}{k}
               };\\
\nonumber C_1&=&-\exp{\!-\frac{2\ro}{k}\!}
          \frac{
                  k\,\,\sh^2\frac{r_{1+}}{k}
               }
               {
                  \sh\frac{r_{1+}-r_{1-}}{k}
               };\quad
          r_{-}=r_{1-}; \quad r_{+}=r_{1+},
\end{eqnarray}
and for the metric(\ref{12})
\begin{eqnarray}
\nonumber A_2&=&\exp{\!-\frac{2\ro}{k}\!}\ch\frac{r_{2-}+r_{2+}}{k};\quad
          B_2=-\exp{\!-\frac{2\ro}{k}\!}
          \frac{
                  k\,\,\sh^2\frac{r_{2-}}{k}
               }
               {
                  \sh\frac{r_{2+}-r_{2-}}{k}
               };\\
\nonumber C_2&=&-\exp{\!-\frac{2\ro}{k}\!}
          \frac{
                  k\,\,\sh^2\frac{r_{2+}}{k}
               }
               {
                  \sh\frac{r_{2+}-r_{2-}}{k}
               };\quad
          r_{-}=r_{2-}; \quad r_{+}=r_{2+}.
\end{eqnarray}
Now we can go to the coordinates in which metrics are regular at
$r_{+}$:
$$ \tilde U=\exp{\!\frac{U}{2C}\!};\quad \tilde V=-\exp{\!-\frac{V}{2C}\!}.$$
Similar coordinates can be constructed to avoid coordinate singularity at
$r_{-}$:
$$ \tilde U=\exp{\!\frac{U}{2B}\!};\quad \tilde V=-\exp{\!-\frac{V}{2B}\!}.$$
The conformal diagram for (\ref{9}) is similar to the
Reissner-Nordstr\"om solution.
Because of a time - like singularity at $r=0$ we cannot continue the
space - time to the negative value of $r$.

The conformal diagram for solution (\ref{12}) differs from the previous one
by absence of  singularities. In principle, the conformal diagram is very
close to the diagram of the Kerr solution (fig.\ref{fig5}).

In domain I the physical 3-space asymptotically turns into the background
when $r \to \infty$. In domain II we may perform
the coordinate transformation similar to (\ref{25}):
\begin{equation}                                             \label{40}
\begin{array}{rclrcl}
      \tilde r&=&-r\,\,\,\exp{-\frac{2\ro}{k}};&\quad
      \tilde t&=&\phantom{-}t\,\,\,\exp{\frac{2\ro}{k}};\\
      \tilde k&=&k\,\,\,\exp{-\frac{2\ro}{k}};&\quad
      \tilde \ro&=&\phantom{-}\ro\,\exp{-\frac{2\ro}{k}};\\
      \tilde r_{2+}&=&-r_{2-}\,\exp{-\frac{2\ro}{k}};&\quad
      \tilde r_{2-}&=&-r_{2+}\,\exp{-\frac{2\ro}{k}}.
\end{array}
\end{equation}
Then the metric (\ref{12}) can be written as
\begin{equation}                                             \label{41}
\d\tilde s^2_{\rm Q2}=\tilde\Lambda_{\rm Q2}\d\tilde t^2
-\frac{1}{\tilde\Lambda_{\rm Q2}}\d\tilde r^2 -
\exp{\!\frac{2\tilde\ro}{\tilde k}\!}\tilde k^2\ch^2\frac{\tilde r}{\tilde k}
\d\Omega^2,
\end{equation}
where
\begin{equation}                                                 \label{42}
\tilde\Lambda_{\rm Q2}=
        \exp{\!-\frac{2\tilde\ro}{\tilde k}\!}
        \frac{
                \sh\left(
                          \frac{\tilde r}{\tilde k}-
                          \frac{\tilde r_{2+}}{\tilde k}
                   \right)
                \sh\left(
                          \frac{\tilde r}{\tilde k}-
                          \frac{\tilde r_{2-}}{\tilde k}
                   \right)
             }
             {
                \ch^2\frac{\tilde r}{\tilde k}
             }.
\end{equation}
It is clear that now $\tilde r_{2+}$ is the event horizon, and
$\tilde r_{2-}$ is the Cauchy horizon. Note that the
asymptotics of the 3-space
in domain II is the Lobachevski space with the curvature radius
$\tilde k \neq k$! Since $r_{2+}$ and $r_{2-}$ have
opposite signs, the throat $r=0$ is spacelike.
If observer goes from domain I into domain III and crosses
the horizon $r_{2+}$, then he must fall through the wormhole, but
he may choose where he appears after that between
two symmetric domains II and II${}^\prime.$ Actually,
we have two enumerable sets of wormholes.

\section{Temperature of the solution with horizons}

The Euclidean section of metric (\ref{4}) is
$$
   \d s^2=\Lambda \d \tau^2+\frac{1}{\Lambda}\d r^2+ R^2\d \Omega^2.
$$
This metric is regular on the horizon \cite{GibHaw}, if the euclidean
time $\tau$ is periodic $0\leq\tau\leq\beta$ with the period
$$\beta=\left.\frac{4\pi}{\frac{\d\phantom{r}}{\d r}\Lambda}\right|_{r=r_+}$$.
The corresponding temperature appears to be
\begin{equation}                                             \label{43}
T=\beta^{-1}=\left.\frac{1}{4\pi}\frac{\d\phantom{r}}{\d
r}\Lambda\right|_{r=r_+}.
\end{equation}

By substituting the corresponding values of $\Lambda$ into this formula
we obtain the temperature $T_{\rm V1}$ (\ref{15}) for the metric (\ref{6}),
the temperature $T_{\rm Q1}$ (\ref{16}) for the metric (\ref{9}) and
the temperature $T_{\rm Q2+}$ (\ref{17}) for the metric (\ref{12}).
We cannot obtain the temperature $T_{\rm Q2-}$ by substituting
 $r_{-}$ instead of $r_{+}$ into (\ref{43}) because in domain III
$r_{-}$ is not the event horizon, it is the Cauchy horizon.
This temperature is observed in domain II, where $r_{2-}$ is the
event horizon. To obtain it, we may substitute the
metric function $\tilde\Lambda_{\rm Q2}$ (\ref{42}) into (\ref{43}).
The resulting expression is (\ref{18}).

The asymptotics are easily calculated by using the follow formulas:
$$
\th\,\Delta_1=\th\left(\frac{\ro}{k}\right)
\sqrt{
\frac{
       1-\frac{Q^2}{k^2}\left(1-\cth{\frac{\ro}{k}}\right)^2
     }
     {
       1-\frac{Q^2}{k^2}\left(1-\th\left(\frac{\ro}{k}\right)\right)^2
     }
},
$$
$$
\th\,\Delta_2=\cth{\frac{\ro}{k}}
\sqrt{
\frac{
       \frac{Q^2}{k^2}\left(1-\th\left(\frac{\ro}{k}\right)\right)^2-1
     }
     {
       \frac{Q^2}{k^2}\left(1-\cth{\frac{\ro}{k}}\right)^2-1
     }
}.
$$
These formulas may be obtained from (\ref{34}, \ref{35}) and
(\ref{38}, \ref{39}).

\section{Conclusion}

Within the context of a spherically symmetric model in the
presence of a nontrivial background we have investigated
some exact solutions of gravitation equations. We have found
that there are two branches of the solutions. The first branch
is close to the corresponding solution of the Einstein equation, and
the second has a set of nontrivial properties
and diverges if the background passes into the Minkovski space.

It is worth noting that the second branches should be of
essentially cosmological origin because the radius
of wormholes is equal to the Universe radius $k$.

In principle, this toy model may be considered as a very
anisotropic cosmological model, in which both local and global
properties are described by the gravitational field produced by the
central-symmetric solution.

The first cosmological model was a static cylinder Universe.
After discovering the cosmological redshift it became
clear that the observed Universe should be nonstationary.

However, in principle, "scattering" of the Galaxies can be understood
in the framework of a static cosmology.

Let us suppose that the Einstein cylinder is actually a throat of
the wormhole with cosmological radius. If the metric is such that
the test bodies are attracted by the throat from one side and repulsed
from another, then it is clear that close to the minimal radius,
the local metric  appears to be nonstationary for a
free falling observer.

Such a model is impossible in the framework of the conventional
General Relativity because the violation of energy conditions
is necessary for the wormhole creation \cite{Wormholes}.
If we consider the gravitation
on the nontrivial cosmological background, similar solutions can
appear.

Much attention has been paid to the investigation
of wormholes. Classical wormholes usually are Euclidean metrics
which consist of two large regions of spacetime connected by a throat.
They play an important role in quantum gravity \cite{HaW1,HaW2}.
It was just the microscopic wormholes which have a Planck size throat.
An analytic continuation of closed cosmological solutions
to imaginary time leads to euclidean wormholes with the size of the throat
equal to the maximum size of the universe \cite{Zhuk}.
Wormholes considered in the present paper demonstrate
similar properties but in the Lorentzian sections.

\newpage
\begin{center} Figure Captions \end{center}

\CAPT{\label{fig1}The metric function and luminosity distance
for the metric (\pref{6}); the dashed curve shows $\Lambda^{-1}$.}

\CAPT{\label{fig2}The metric function and luminosity distance
for the metric (\pref{7}).}

\CAPT{\label{fig3}Horizons of the metric (\pref{12}).}

\CAPT{\label{fig4}A minimal possible
value of the horizon for metric (\pref{12}).}

\CAPT{\label{fig5}Conformal diagram for the solution (\pref{12}).}

\CAPT{\label{fig6}Temperatures of the metrics
(\pref{6}) and (\pref{9}); $r_{\rm 0 min}=k\,\arcth\frac{Q}{k-Q}$.}

\CAPT{\label{fig7}Spatial sections of the background metric embedded
into a fictitious high - dimensional space.}

\CAPT{\label{fig8}Conditional representation of the space - time of the
wormhole.}

\CAPT{\label{fig9}Spatial sections of the physical metric. The background
metric
is also shown. Continuous line in the background space
corresponds to the discontinuity of the physical line at $r=0$.}

\CAPT{\label{fig10}The metric function for the metric(\pref{9}) for different
values of $Q$.}

\CAPT{\label{fig11}The metric function for the metric(\pref{12}) for different
values of $Q$.}

\end{sloppypar}
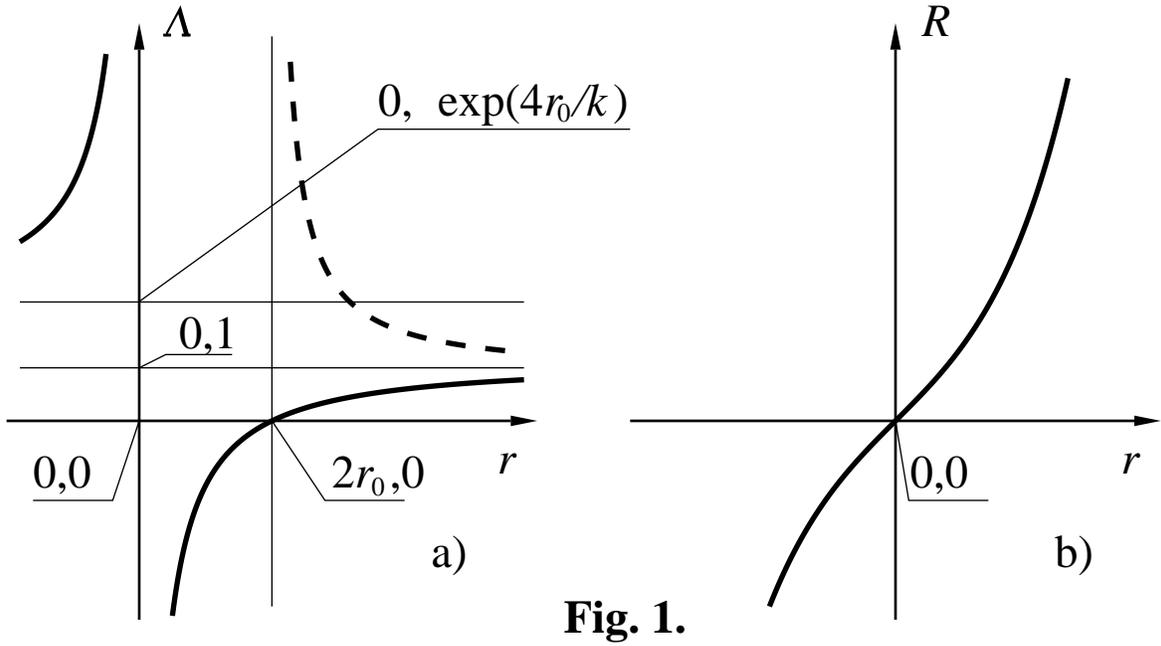
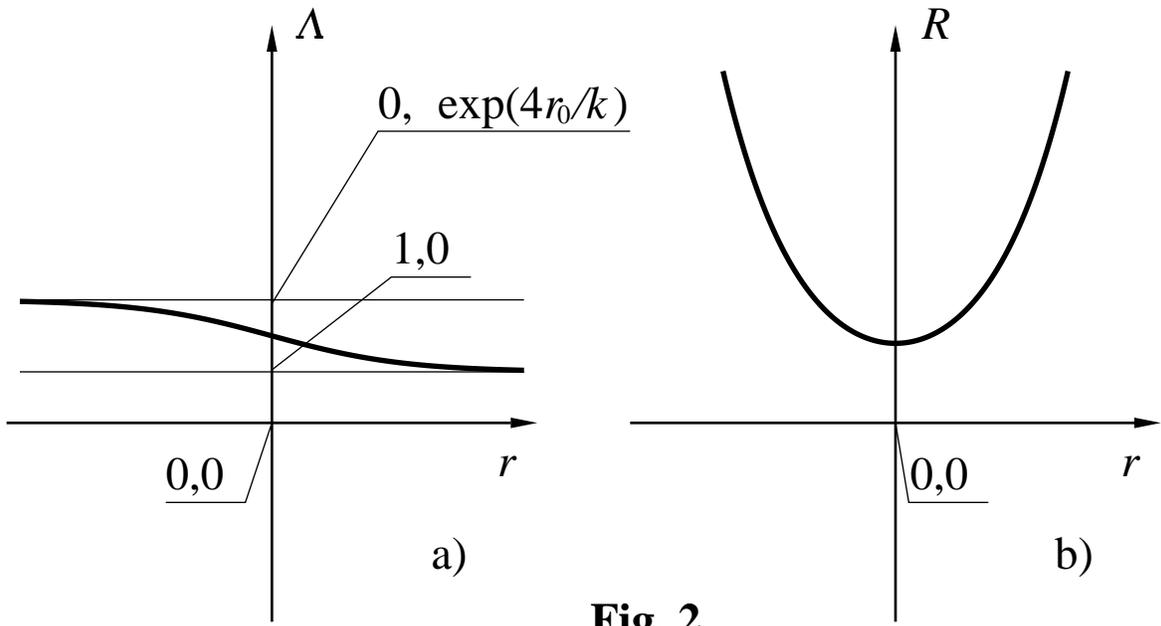
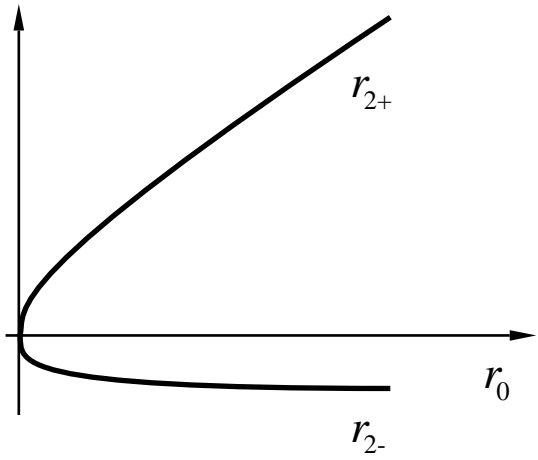
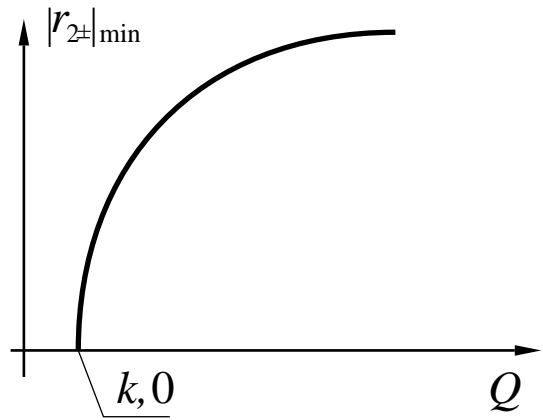
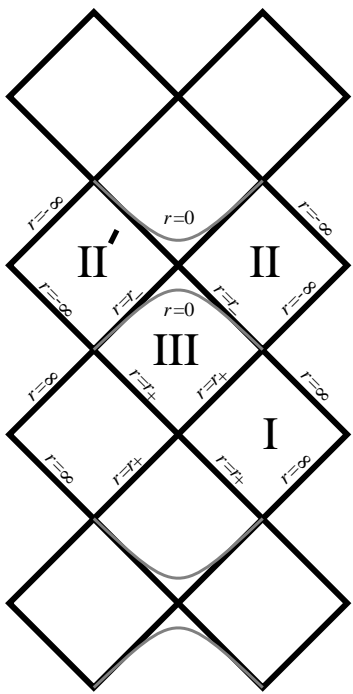
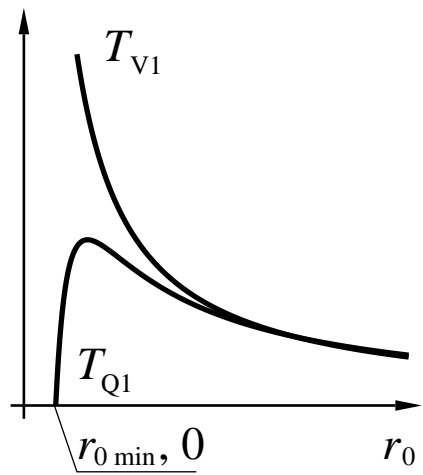
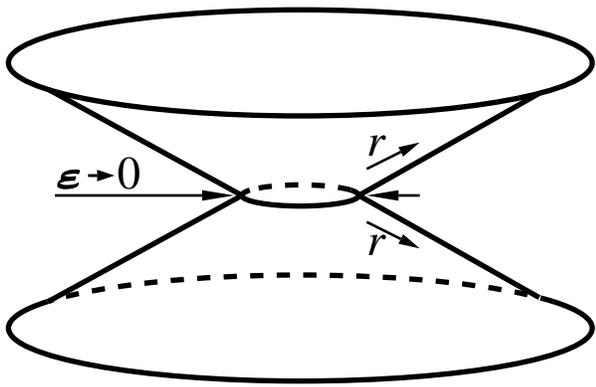
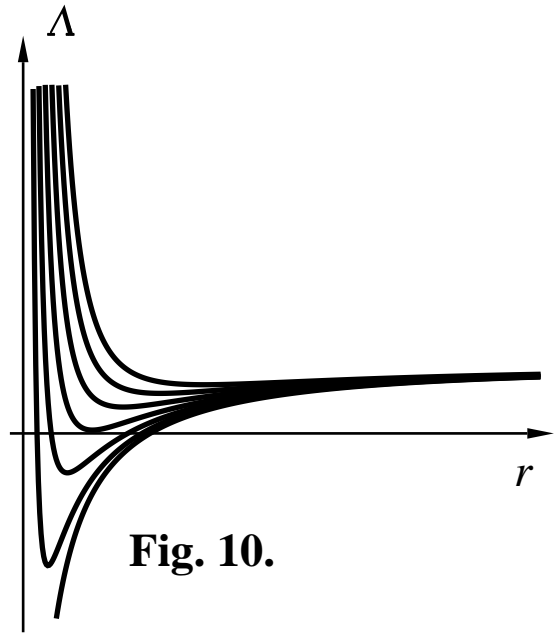
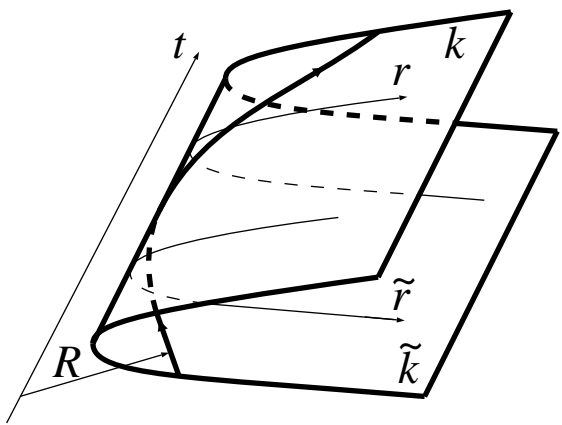
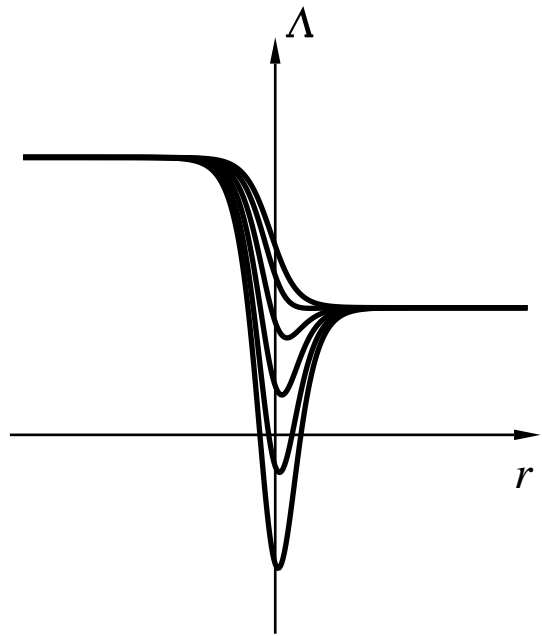
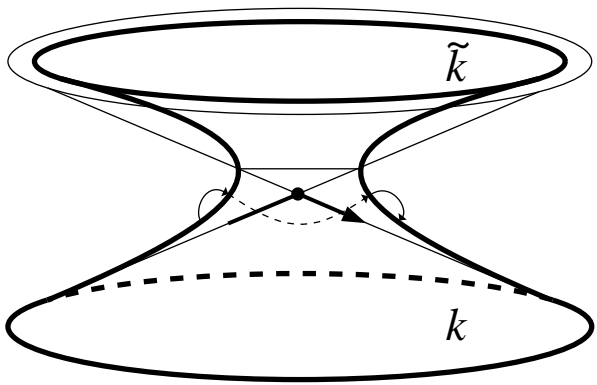
\end{document}